\documentclass[11pt,preprint,natbib]{aastex}

\begin{document}
\newcommand{\be}{\begin{equation}}
\newcommand{\ee}{\end{equation}}
\newcommand{\bea}{\begin{eqnarray}}
\newcommand{\eea}{\end{eqnarray}}
\newcommand{\ba}{\begin{array}}
\newcommand{\ea}{\end{array}}

\title{On the Multiple Deaths of Whitehead's Theory of Gravity} 

\author{Gary Gibbons}
\affil{Department of Applied Mathematics and Theoretical Physics,\\
Cambridge University, UK} 

\and

\author{Clifford M. Will}
\affil{McDonnell Center for the Space Sciences, Department of Physics \\
Washington University, St. Louis, USA\footnote{Permanent address}, and \\
GReCO - Gravitation et Cosmologie, 
Institut d'Astrophysique, Paris, France}


\begin{abstract}
Whitehead's 1922 theory of gravitation continues to attract the attention of
philosophers, despite evidence presented in 1971 that it violates
experiment.  We demonstrate that the theory strongly fails
five quite different experimental tests, and conclude that, notwithstanding
its meritorious philosophical underpinnings, Whitehead's theory is truly
dead.
\end{abstract}
 
\section{Introduction and summary}
\label{sec:intro}

In 1922, the distinguished mathematician and philosopher Alfred North
Whitehead (1861-1947) ,
then in his 60th   year, published a relativistic  theory of gravity
with the property, which it shares with  Einstein's
theory,  of containing no arbitrary parameters. 
Furthermore, when suitably interpreted,  it  yields the same 
predictions as General Relativity (GR), not only for the three classic tests of
light bending, gravitational redshift and the precession
of the perihelion of Mercury, but also for
the Shapiro time delay effect \citep{Shapiro}, recently confirmed
to one part in $10^5$ \citep{Bertotti}.

The reason for this coincidence  was realized early on 
by  \cite{Eddington}. In the case
of vanishing cosmological constant the Schwarzschild solution
is not only an exact solution of  Einstein's theory, it
is an exact solution of Whitehead's theory as well.
Thus it gives the same predictions for the parametrized post-Newtonian
(PPN) parameters
$\gamma=\beta=1$. Eddington's remark nicely explained
an observation of \cite{Temple2} that 
the predictions of the precession of the perihelion
for the two theories agree exactly, and gave 
rise to the (incorrect) idea that it  is
indistinguishable from GR,  a point refuted by \cite{Harvey}
by the observation that Birkhoff's theorem fails for Whitehead's
theory: the field outside a spherically symmetric source is
not just given by the Schwarzschild solution but in general
contains an additional constant of integration which is
in principle measurable. 

In fact an even stronger statement can be made. This remarkable
correspondence of exact solutions extends to the Kerr solution
\citep{RussellWasserman} and
thus to  the
corresponding Lense-Thirring or frame dragging effects \citep{Rayner2}.
Thus experiments such as that involving  the
LAGEOS satellites \citep{CiufoliniPavlis} which
have verified the effect at the 10-15\% level and the ongoing
NASA-Stanford  Gravity Probe B
superconducting gyroscope experiment, which aspires to an accuracy of 1\%,
cannot distinguish 
Whitehead's from Einstein's theory on the basis of frame dragging (we will
see below that LAGEOS actually 
tests Whitehead because of the failure Birkhoff's
theorem).  

The mathematical explanation for  this striking, but accidental,
coincidence
is that both the Schwarzschild solution and the Kerr solution may be cast
in Kerr-Schild form \citep{KerrSchild}.
That is, coordinates exist for which   
\be
g_{\mu \nu} = \eta_{\mu \nu} + l_\mu l_\nu ,
\label{kerrschild}
\ee
where 
\be
\eta^ {\mu \nu} l_\mu l_\nu = g^{\mu \nu} l_\mu l_\nu =0\,. \label{eqn1}
\ee
and $l_\mu $ is tangent to a null geodesic congruence,
\be
l_{[\mu ; \nu } l ^ \nu  l _{\lambda]}  =0\,, \label{eqn2}
\ee  
where $l^\mu$ is obtained from $ l_\mu$
by index raising using either the metric $\eta_{\mu \nu}$ or the  metric
$g_{\mu \nu}$.
It follows \citep{KerrSchild}   
that 
\be
 h_{\mu \nu}=   l_\mu l_\nu \label{lin} 
\ee
satisfies
the {\sl linearized} Einstein equations. If this can be chosen to
agree with Whitehead's retarded solution, then his metric and that
of Einstein will agree exactly.  

Thus for a single particle at rest at the origin,  
in spherical polar Minkowski  coordinates $t,r, \theta ,\phi$,
Whitehead's metric is 
\be
ds ^2 = -dt^2 + dr ^2 + r^2 ( d \theta ^2 + \sin ^2 \theta  d \phi ^2 ) 
+ { 2 M \over r} (dt -dr ) ^2.
\ee
On the other hand, the Schwarzschild metric, 
in standard Schwarzschild coordinates 
$(T,r,\theta, \phi ) $ is
\be
ds ^2 = -(1-{ 2M \over r} ) dT^2 + 
{ dr ^2 \over 1-{2 M \over r} }   + 
r^2 ( d \theta ^2 + \sin ^2 \theta d \phi ^2 ) \, .  
\ee
If we set
\be
t= T- M \ln ( {r \over 2  M} -1)\,,  
\label{eddingtontransform}
\ee
the two metrics are seen to coincide, which is Eddington's observation
\citep{Eddington}.
If we define $u=t-r$, then the coordinates $u, r, \theta ,\phi$ are
nowadays referred to as outgoing  Eddington-Finkelstein coordinates.
Thus Whitehead's spacetime manifold is geodesically incomplete
\label{hmunu}
with respect to his  curved metric because outgoing   Eddington-Finkelstein
coordinates  cover only the lower half of the full Kruskal manifold.
The surface $ r=2 M$ is the past event horizon, and Whitehead's
particle is naked and
corresponds  to what is now called a White Hole \citep{Harvey},
the time reverse
of a Black Hole. The Kerr solution is also of Kerr-Schild form
and is also an exact solution of Whitehead's metric \citep{RussellWasserman}
when expressed in terms of advanced null coordinates.
It corresponds therefore to a naked rotating White Hole.
For strong fields therefore, even if a single object is considered,
the two theories would be expected to make very different predictions.
Note also that this exact correspondence between solutions of
Whitehead's theory and solutions of Einstein's theory
holds only for a special class of solutions. Not every solution
$h_{\mu \nu}$ of linear theory may be cast in the form (\ref{lin}) such that
(\ref{eqn1}) and (\ref{eqn2}) 
hold, i.e to be of Kerr-Schild form. Moreover not every solution
in the  Kerr-Schild class need  be of the retarded form  specified by 
Whitehead. Thus there is no  general agreement between the predictions
of Einstein and those of Whitehead.

In any case, for many years the two  theories were considered to be 
experimentally indistinguishable, and this gave rise to
much philosophical discussion as to whether additional
criteria, for example aesthetic considerations or philosophical
preconceptions,  were needed  in order to reject or accept
one of them.  This is brought out 
in  Broad's review of Whitehead's book {\it The Principle of
Relativity} \citep{Broad}, and
a particularly clear discussion 
indicative of the mood in the late 1950's is that of 
\cite{Bonnor}.

From today's perspective, one can say that
the principal difference between Einstein and Whitehead is
the latter's insistence on fixed {\it a priori} 
spatio-temporal relations, which in practice meant the adoption
of a fixed, and in particular  unobservable, 
background Minkowski spacetime. This is  stated with admirable  clarity
by the philosopher
John \cite{Bain}, who provides a valuable account of how
Whitehead's ideas about relativity were embedded in his overall 
philosophy of nature (see also  \cite{Tanaka}).   

In fact, by the late 1960's the promise of new technology
had led to a more optimistic, empirical  viewpoint, and  
rival  theories of gravity were carefully scrutinised
both for internal consistency and for testable  
predictions additional to the three classic tests.
An important milestone was Shapiro's time delay
prediction \citep{Shapiro}. An outcome of this  
line of research was the discrediting by 
\cite{Will}
of Whitehead's theory.  

However,
Whitehead's philosophical ideas continue to attract widespread attention,
often under the rubric of Process Philosophy, 
and perhaps because of his formidable achievements in 
the foundations of mathematics and logic.
He was after all co-author with Bertrand Arthur William Russell 
(1872-1970)  
of the epoch making {\it Principia
Mathematica}.  As a result, many
of his followers have been reluctant to abandon his
theory of gravity despite the growing observational  evidence against it.

Will's original disproof  of Whitehead's theory  was  based on the fact
that Whitehead's theory predicts an anisotropy in the ``locally measured''
Newton's constant
due to distant matter. Thus a mass $M$ at a distance $r$
from the Earth produces an effective  Newton's contant
\be
G_{eff}= G \left ( 1+  \frac{2 G M }{ rc^2} + \frac{GM}{ rc^2} \cos^2 \theta  
\right )\,. 
\label{anisotropy}
\ee 
where $\theta$ is the angle between the Earth's radial direction
and the distant gravitating body.  This would 
produce anomalous Earth tides that
would show up in gravimeter 
experiments, yet there was no experiment evidence for such effects
\citep{goodkind}.
As a critique of Will's argument, 
it was pointed out that the resultant Earth tides depend
on the distribution of extra-solar system  matter \citep{Mentock}
whose distribution is uncertain, and so a cancellation might take place.
However, as we shall show below,
allowing for these uncertainties will not change the predicted effect
sufficiently  to invalidate  Will's  argument.
Another attempt to avoid  Will's argument was to change
the interpretation \citep{ReinhardtRosenblum, Hyman}.
In \cite{ChiangHamity} it was shown that the re-interpretation
of \cite{ReinhardtRosenblum} would not achieve this goal,
and they obtained the same result for the anisotropy of Newton's
constant (\ref{anisotropy}) as did  Will. These  general 
conclusions,  while  accepted by \cite{Bain}, were rejected 
by  
\cite{Fowler} and  the latter's remarks were reiterated by \cite{Tanaka}.
Similar reservations  have been expressed by 
\cite{RussellWasserman}.   

In fact, normally in science a single incorrect prediction
is regarded as sufficient grounds for rejecting a theory;
hence the well known dictum of   ``Darwin's Bulldog" 
\begin{quote}
    The great tragedy of Science - the slaying of a beautiful hypothesis 
by an ugly fact.
        
Thomas H. Huxley (1825 - 1895)

\end{quote} 

By contrast, as we are reminded by \cite{Popper}, 
the confirmation of a theory is never complete. 
The best one can do is to subject it to increasingly  precise and
exacting tests covering  a wider and wider range of phenomenona
and circumstances. 

It turns out that Whitehead's theory is definitely excluded
by several modern experiments, and our aim in this
article
and the reason for our title is to point out that any
one of them is sufficient for
rejection.  In other words judged by modern scientific and
technological
standards, Whitehead's theory, beautiful as it may seem in the eyes
of many of its beholders, is truly dead.
By contrast, Einstein's theory passes all of 
these tests with flying colors.

Specifically, Whitehead's theory fails five tests, most of them by many
orders of magnitude

\begin{enumerate}

\item
{\it Anisotropy in G}. We have reexamined Will's 1971 derivation,
incorporating a model for the mass distribution of the galaxy that includes a
dark matter halo.  The predicted effect is still at least 100 times larger
than the experimental bound.

\item
{\it Nordtvedt effect and lunar laser ranging}.  
Whitehead's theory predicts that massive, self-gravitating bodies violate
the weak equivalence principle in that their acceleration in an external
gravitational field depends on their gravitational binding energy (Nordtvedt
effect).  The predicted size is 400 times larger than that permitted by
lunar laser ranging.

\item
{\it Gravitational radiation reaction and the binary pulsar}.  The theory
predicts anti-damping of binary orbits due to gravitational radiation
reaction at a level $(v/c)^3$ beyond Newtonian gravity, in contrast to the
$(v/c)^5$ damping effect in GR.  Thus it strongly violates binary pulsar
data by about four orders of magnitude, and with the wrong sign.

\item
{\it Violation of Birkhoff's theorem, and LAGEOS satellites}.  The static,
spherically symmetric solution of the theory for finite sized bodies has an
additional contribution dependent on the body's size 
\citep{Harvey,Rayner1,Synge1}.
This produces an additional advance of the perigee of the LAGEOS II
satellite, in disagreement with observations by a factor of 10.

\item
{\it Momentum conservation and the binary pulsar}.  Whitehead's
theory predicts an acceleration of the center of mass of a binary system, a
violation of momentum conservation \citep{Clark}.  Precise timing of the
pulsar B1913+16 in the Hulse-Taylor binary pulsar rules out this effect 
by a factor of a million.

\end{enumerate}

Any of these tests alone would have been enough to kill Whitehead's theory,
so collectively they amount to overkill.  On the other hand they illustrate
both the precision and depth that modern technology has brought to the
problem of testing gravity, and serve as a warning to any would-be inventor
of an alternative gravity theory, or to anyone who might hope that a
suitably modified or reinterpreted Whiteheadian theory would pass muster
\citep{Schild1,Hyman,ReinhardtRosenblum}.
It is not sufficient to check the ``classic tests'' of light bending,
perihelion advance of Mercury,
and gravitational redshift.  There is now an exhaustive
battery of empirical checks that must be done.  

The remainder of this paper provides some technical details to support these
conclusions.  Throughout, we adopt the ``canonical'' version of Whitehead's
theory, 
specified as follows.  One first
assumes the presence of a
flat background metric $\eta_{\mu\nu}$, whose Riemann tensor vanishes
everywhere.  This background metric defines null cones for any chosen
spacetime event $x^\mu$, given by points $x^{\prime\mu}$ satisfying 
\be
\eta_{\mu\nu} y^\mu y^\nu = 0 \,, \quad y^\mu = x^\mu - x^{\prime\mu}
\,.
\ee
The physical metric $g_{\mu\nu}$ is then given by (henceforth
we use units in which $G=c=1$)
\bea
g_{\mu\nu}(x^\alpha) &\equiv& \eta_{\mu\nu} - 2 \sum_a m_a \frac{(y_a^-)_\mu
(y_a^-)_\nu}{(w_a^-)^3} \,, \nonumber \\
(y_a^-)^\mu &=& x^\mu - (x_a^-)^\mu \,,\nonumber  \\
\eta_{\mu\nu} (y_a^-)^\mu (y_a^-)^\nu &=&0 \,, \nonumber \\
w_a^- &=& \eta_{\mu\nu}(y_a^-)^\mu (dx_a^\nu /d\sigma)^- \,, \nonumber \\
d\sigma^2 &=& \eta_{\mu\nu} dx^\mu dx^\nu \,,
\eea
where the sum is over over all particles, with rest mass
$m_a$.
Indices on $(y_a)^\mu$ are raised and lowered using
$\eta_{\mu\nu}$.  The quantities $(x_a^-)^\mu$, $(dx_a^\nu /d\sigma)^-$ 
are to be
evaluated along the past flat null cone of the field point $x^\mu$.

Following \cite{Synge1}, we assume that matter fields couple
{\it only} to the physical metric $g_{\mu\nu}$.  
This makes Whitehead's theory a ``metric theory'' of gravity (see
\cite{tegp}
for discussion).  As such, it automatically
satisfies the Einstein Equivalence Principle (EEP), which has been
verified to extremely high precision using laboratory
E\"otv\"os-type experiments (parts in $10^{13}$) 
and gravitational redshift experiments (parts in $10^4$), among
others.
The background metric then
has no further direct physical consequences, apart from 
its role in defining the
physical metric.
This point mass expression can be generalized to continuous fluids in a
straightforward way \citep{tegp}.

\section{Post-Newtonian limit and gravitational radiation reaction in
Whitehead's theory}
\label{sec:PNlimit}

\subsection{Solution of Whitehead's theory to 1.5 post-Newtonian order}
\label{sec:15PN}

We wish to evaluate the Whitehead metric within the near-zone
of a slow-motion
gravitating system, in order to derive the equations of motion.  This
corresponds to field points such that $|{\bf x}| \ll \lambda \sim R/v$,
where $\lambda$ is roughly a gravitational wavelength, 
$R$ and $v \ll 1$ are the
characteristic size and internal velocity of the system.
Accordingly, we want to evaluate $g_{\mu\nu}$ at $(t,{\bf x})$ in terms of
source varables ${\bf x}_a$ evaluated at the {\it same} time $t$.
We make the standard assumption of post-Newtonian theory that $v^2 \sim m/r
\sim \epsilon$, where $\epsilon$ is a small parameter used for bookeeping
purposes.  Our goal is to determine the metric through 1.5
post-Newtonian order, or to order $\epsilon^{3/2}$ beyond Newtonian gravity;
this involves evaluating $g_{00}$ through $O(\epsilon^{5/2})$,
$g_{0j}$ through $O(\epsilon^{2})$,
and $g_{ij}$ through $O(\epsilon^{3/2})$.  
This will include the usual post-Newtonian terms
relevant for solar-system tests, as well as, it will turn out, the leading
effects of gravitational radiation reaction in this theory.  

We expand the {\it retarded} position of the $a$-th particle by
\bea
{\bf x}_a^- &\equiv& {\bf x}_a (t-|{\bf x}-{\bf x}_a^-|) \nonumber \\
&\approx& {\bf x}_a  - {\bf v}_a |{\bf x}-{\bf x}_a^-|
+ \frac{1}{2} {\bf a}_a |{\bf x}-{\bf x}_a^-|^2 + \dots
\,,
\eea
where ${\bf x}_a$, ${\bf v}_a$ and ${\bf a}_a$ are the position, velocity
and
acceleration of the $a$-th
particle at the field-point time $t$.  
We can then expand the spatial component $(y_a^-)^i= (x-x_a^-)^i$
in terms of the instantaneous difference $z_a^i \equiv (x-x_a)^i$ 
according to 
\be
(y_a^-)^i= z_a^i + \epsilon^{1/2} v_a^i y - \frac{1}{2} \epsilon a_a^i y^2
+ \frac{1}{6} \epsilon^{3/2} {\dot a}_a^i y^3 + O(\epsilon^2) \,,
\ee
where $y \equiv |{\bf y}_a^-|$.
We also expand the retarded velocity component $(v_a^-)^i \equiv
(dx_a^i /dt)^-$ according to
\be
(v_a^-)^i = v_a^i - \epsilon^{1/2} a_a^i y +\frac{1}{2} \epsilon {\dot
a}_a^i y^2 + O(\epsilon^{3/2}) \,.
\ee
Note that, because the quantity $(y_a^-)^\mu$ is null with respect to the
flat metric, $(y_a^-)^0 = y$, and thus 
\be
w_a^- = (dt/d\sigma)^- (-y + \epsilon^{1/2} {\bf y}_a^- \cdot {\bf v}_a^- )
\,.
\ee
The foregoing expressions can then be iterated to the required order in
$\epsilon$ to convert all expressions into functions of ${\bf v}_a$, ${\bf
a}_a$, ${\dot {\bf a}}_a$, ${\bf z}_a$, and $z_a = |{\bf z_a}|$.  The result is 
\be
g_{\mu\nu} = \eta_{\mu\nu} + \epsilon \sum_a m_a (h_a)_{\mu\nu} \,,
\ee
where
\bea
(h_a)_{00} &=& \frac{2}{z_a} + \epsilon^{1/2} \frac{4 {\bf v}_a \cdot {\bf
z}_a}{z_a^2} + \epsilon \left [ 2 \frac{ v_a^2}{z_a} -5 \frac{{\bf a}_a \cdot
{\bf z}_a}{z_a} + \frac{({\bf v}_a \cdot {\bf z}_a)^2}{z_a^3} \right ] 
\nonumber \\
&& +  \epsilon^{3/2} \left [\frac{8}{3} \dot{\bf a}_a \cdot {\bf z}_a
 - 2 {\bf v}_a \cdot {\bf a}_a + 6 \frac{v_a^2 {\bf v}_a \cdot {\bf z}_a}{z_a^2}
 -12 \frac{{\bf v}_a \cdot {\bf z}_a {\bf a}_a\cdot {\bf z}_a}{z_a^2}
 -4\frac{({\bf v}_a \cdot {\bf z}_a)^3}{z_a^4} \right ] +O(\epsilon^{2})\,, 
 \nonumber \\
(h_a)_{0j} &=& -\frac{2z_a^j}{z_a^2} - \epsilon^{1/2} \left [ 
2 \frac{v_a^j}{z_a} + 2\frac{{\bf v}_a \cdot {\bf z}_a z_a^j}{z_a^3} \right ]
\nonumber \\
&& + \epsilon \left [ a_a^j + 4 \frac{{\bf a}_a \cdot {\bf z}_a z_a^j}{z_a^2}
- \frac{v_a^2 z_a^j}{z_a^2} + 2 \frac{({\bf v}_a \cdot {\bf z}_a)^2 z_a^j}{z_a^4}
- 4 \frac{{\bf v}_a \cdot {\bf z}_a v_a^j}{z_a^2} \right ] 
+ O( \epsilon^{3/2}) \,,  \nonumber \\
(h_a)_{ij} &=&  \frac{2z_a^i z_a^j}{z_a^3} + \epsilon^{1/2} \frac{4z_a^{(i}
v_a^{j)}}{z_a^2} + O(\epsilon)  \,.
\label{hsuba}
\eea
Indices on spatial vectors are raised and lowered using the Cartesian
metric; parentheses around indices denote symmetrization, while
square brackets denote antisymmetrization.

The first term in $(h_a)_{00}$ can be recognized as yielding
the normal Newtonian
potential $U$, given by
\be
U(t,{\bf x}) = \sum_a \frac{m_a}{z_a} = 
\sum_a \frac{m_a}{|{\bf x} - {\bf x}_a|} \,.
\label{newt}
\ee
Note the presence of 0.5PN terms in the metric; these are terms of order 
$\epsilon^{1/2}$ in $(h_a)_{00}$, and
$\epsilon^{0}$ in $(h_a)_{0j}$.
Because of general covariance, we are free to change coordinates to
manipulate the form of the physical metric.  In particular, we can remove these 0.5PN
terms, can manipulate the PN terms to put them into a form to make
comparisons with the standard parametrized post-Newtonian (PPN)
framework \citep{tegp}, and can
simplify the 1.5PN terms.  Even though the background metric
$\eta_{\mu\nu}$ will change its form under such coordinate transformations,
this will have no physical consequences, since only $g_{\mu\nu}$ couples to
matter.

The following coordinate transformation kills the 0.5PN terms in the
physical metric, puts the PN terms into the standard PPN gauge, and also
kills the 1.5PN terms in $(h_a)_{ij}$:
\bea
t &=& {\bar t} - 2 \epsilon {\bar L}^0 + 
\frac{5}{2} \epsilon^{3/2} {\bar X}_{,{\bar 0}} + O(\epsilon^{2}) \,, \nonumber
\\
x^i &=& {\bar x}^i 
+ \epsilon {\bar X}_{,{\bar j}} 
- 2 \epsilon^{3/2} {\bar L}^j + O(\epsilon^{2}) \,,
\label{transform1}
\eea
where commas denote partial derivatives, and where
\bea
{\bar L}^0 &=& \sum_a m_a \ln {\bar z}_a \,, \nonumber \\
{\bar L}^j &=& \sum_a m_a {\bar v}_a^j \ln {\bar z}_a \,, \nonumber \\
{\bar X}&=& \sum_a m_a  {\bar z}_a \,.
\label{LX}
\eea
Note that the first term in the time transformation is the post-Newtonian
analogue of the \cite{Eddington} transformation.
In carrying out 
the normal coordinate transformation,
\be
g_{{\bar \alpha}{\bar \beta}}({\bar x}^{\bar \gamma})
= \frac{\partial x^\mu}{\partial {\bar x}^{\bar \alpha}}
 \frac{\partial x^\nu}{\partial {\bar x}^{\bar \beta}}
g_{\mu\nu}(x^\lambda) \,,
\ee
to 1.5PN order, it is also necessary
to reexpress the potentials in terms of the new
coordinates.  For example, the Newtonian potential changes according to
\be
U \to U - \epsilon (U^2 + \Phi_2 + \Phi_W)
- 2\epsilon^{3/2} [ L^0 {\dot U} + L^j U_{,j} - \Sigma (L^j)_{,j}
 + \Sigma (v^j L^0)_{,j} ] \,,
\ee
where all quantities on the right side are barred, and where
\be
\Sigma (f) \equiv \sum_a \frac{m_a f(t,{\bf x}_a)}{z_a} \,.
\ee
The potentials $\Phi_2$ and $\Phi_W$ are defined below.  

A further coordinate transformation, given by
\be
{\bar t} = t^\prime - \epsilon^2 
( 4 U^\prime L^{\prime 0} + 2 X^\prime_{,j} L^{\prime 0}_{,j}
                -2 M^{\prime j}_{,j}) \,,
\label{transform2}
\ee
where $M^j = \sum_a m_a X_{,j}({\bf x}_a) \ln z_a $, 
simplifies the 
1.5PN terms in $g_{0j}$ and $g_{00}$.

The post-Newtonian part of the metric will be discussed in Sec.
\ref{sec:PPN}, while the 1.5PN part will be discussed in Sec. 
\ref{sec:reaction}.

\subsection{PPN Parameters}
\label{sec:PPN}

Following the coordinate transformation of Eqs. (\ref{transform1}), 
the metric to PN
order takes the form
\bea
g_{00} &=& -1 + 2U  -2U^2 - 3 \Phi_1 - 2\Phi_2  + 6
{\cal A} - 2 \Phi_W  \,, \nonumber
\\
g_{0j} &=&  -4 V^j - \frac{7}{2} W^j \,, \nonumber
\\
g_{ij} &=& \delta_{ij} (1+2U) \,,
\eea
where we drop the explicit use of $\epsilon$, and 
where the potentials are given by
\bea
\Phi_1 &=& \sum_a \frac{m_a v_a^2}{z_a} \,, \quad
\Phi_2 =  \sum_a \frac{m_a U({\bf x}_a)}{z_a} = \sum_{a,b\ne a} 
\frac{m_a m_b}{z_a z_{ab}}\,, \nonumber \\
{\cal A} &=& \sum_a \frac{m_a ({\bf v}_a \cdot {\bf z}_a)^2}{z_a^3} \,, \quad
\Phi_W = \sum_{a,b \ne a} \frac{m_a m_b {\bf z}_a}{z_a^3} \cdot
\left ( \frac{{\bf z}_{ab}}{z_b} - \frac{{\bf z}_{b}}{z_{ab}} \right )  \,,
\nonumber \\
V^j &=& \sum_a \frac{m_a v_a^j}{z_a} \,, \quad
W^j = \sum_a \frac{m_a {\bf v}_a \cdot {\bf z}_a z_a^j}{z_a^3} \,,
\eea
where ${\bf z}_{ab} = {\bf z}_a -{\bf z}_b$.

We now compare this metric with the PPN metric for
point masses given in \cite{tegp,livrev06}.  
Transforming from the perfect fluid version
of $U$ to the point mass version using $U_{\rm fluid} = U_{\rm point} -
\frac{1}{2} \Phi_1 - 3 \gamma \Phi_2$, and working in the universal
rest frame, it is a simple matter to read off the PPN parameters,
\bea
\gamma &=& 1 \,, \quad \beta = 1 \,, \quad \xi = 1 \,, \nonumber \\
\alpha_1 &=& 0 \,, \quad  \alpha_2 = 0 \,, \quad \alpha_3 =0 \,, \nonumber
\\
\zeta_1 &=& -4 \,, \quad \zeta_2 = -1 \,.
\label{ppnvalues}
\eea
Because we are dealing with point masses rather than perfect fluids, the
PPN parameters $\zeta_3$ and $\zeta_4$, associated with energy density and
pressure are not determined.  

The parameters $\gamma$ and $\beta$ are the same as in GR.  The potential
$\Phi_W$ is the infamous ``Whitehead'' potential, which did not appear in
earlier versions of the PPN framework 
\citep{nordtvedt2,will71,willnordtvedt72,varenna}.
It was later seen to be a generic consequence of any ``quasi-linear''
theory of gravity \citep{will73}.  The original PPN framework was then
extended to incorporate naturally
this potential with its associated ``Whitehead
parameter'', $\xi$.  In Whitehead's theory, $\xi=1$, while in GR, $\xi=0$.
The parameters $\alpha_i$ all vanish, as they do in GR, indicating that
the theory satisfies a kind of Lorentz invariance for gravity, and has no
``preferred-frame'' effects.  This is not surprising, given that it 
is constructed using a flat background Minkowski
metric.  Here we ignore any coupling
between local gravity and a background cosmological solution for the metric,
which can in fact lead to non-zero $\alpha$'s, even with a flat 
background metric
(see \cite{lee76} for an example).  
The ``conservation-law'' parameters $\zeta_i$ are
non-zero, indicating that the theory lacks global conservation laws for
momentum and angular momentum; in GR, the $\zeta$'s all vanish.  In Sec.
\ref{sec:tests} we will see that many of these values are in violation of
experimental bounds.

\subsection{Static spherically symmetric metric}
\label{sec:SSSmetric}

For a single, static point mass $M$, the metric is particularly simple.  Placing
the mass at the origin of coordinates, we see that $(y_a^-)^0 = y = r$,
$(y_a^-)^j = x^j$, and $w_a^- = y = r$.  The metric then is 
$g_{00} = -1 + 2M/r$, $g_{0j} = -2Mx^j/r^2$, $g_{ij}=\delta_{ij} + 2Mx^i
x^j/r^3$.  The coordinate transformation (\ref{eddingtontransform}) converts
the metric to the Schwarzschild metric of GR \citep{Eddington}.  
This was the basis of the
claim made in the early years of Whitehead's theory that it
satisfied all the ``classic'' tests.  

However, real bodies such as the Sun and Earth are not point masses, but
are finite sized objects made up of many masses.  Working in the PN limit and
assuming a spherically
symmetric collection of masses centered at the origin, it is easy to show
that, for a field point outside the body,
\be
U = M/r \,, \quad X = Mr + I/3r \,, \quad \Phi_W = -\Phi_2 - MI/3r^4 \,,
\ee
where the latter follows from manipulating the identity 
$\nabla^2 (\Phi_W +2U^2 - 3\Phi_2) = -2 X_{,ij} U_{,ij}$, and where 
$M = \sum_a m_a$ and $I = \sum_a m_a r_a^2$ are the total mass and spherical
moment of inertia of the body.  
All other post-Newtonian potentials vanish.
Thus, in the PPN framework, the metric for a finite
spherically symmetric static body becomes
\citep{Harvey,Synge1}
\bea
g_{00} &=& -1 + 2M/r - 2\beta (M/r)^2 + 2 \xi MI/3r^4 \,, \nonumber \\
g_{0j} &=& 0 \,, \nonumber \\
g_{ij}&=& \delta_{ij} (1+2\gamma M/r) \,.
\eea
Recall that $\gamma=\beta=\xi=1$ in Whitehead's theory.  The  
pericenter advance per orbit of a test particle moving on a geodesic of this
metric is given by
\bea
\Delta \omega &=& \frac{6 \pi m}{p}  \left [ (2 +2\gamma -\beta) +
\frac{2\xi}{3} \frac{I}{mp^2} (1+\frac{1}{4}e^2) \right ] \,,
\nonumber \\
&=& \frac{6 \pi m}{p}  \left [1 + \frac{2}{3} \frac{I}{mp^2}
(1+\frac{1}{4}e^2) \right ] \,,
\label{pericenter}
\eea
where $p=a(1-e^2)$, with $a$ and $e$ being the semi-major axis and
eccentricity of the orbit, and where the second line is the Whitehead
prediction.  
The size-dependent term in $\Delta \omega$ 
has a negligible effect on the
perihelion advance of Mercury, and so Whitehead's theory agrees with
the data; however that term {\it will} have measurable 
consequences for ranging of
the Earth-orbiting LAGEOS satellites (Sec. \ref{sec:lageos}).

\subsection{Anisotropy in the locally measured $G$}
\label{sec:Ganisotropy}

Although we have set the fundamental coupling constant $G$ equal to unity by
making a specific choice of units, it turns out that, in many alternative
theories, the ``locally measured'' $G$ may vary.  By locally measured $G$ we
mean the output of a Cavendish-type experiment, whereby one measures the
force between a test body and a source body separated by a chosen distance.  
The result may depend on the velocity of the laboratory relative to a
preferred frame, if any of the $\alpha$ PPN parameters is non-zero, and may
also depend on the presence of matter outside the laboratory.
In the case of Whitehead's theory, there are no preferred-frame effects, but
there are ``preferred location'' effects.  Specifically (see Eq. (6.75) of
\cite{tegp}), the locally measured $G$ is given by
\be
G_{\rm local} = 1 + \frac{7}{3} U_{\rm ext} + \left ( 1-\frac{3I}{MR^2}
\right ) {\hat e}^i {\hat e}^j U_{\rm ext}^{<ij>} \,,
\label{anisotropy2}
\ee
where $I$, $M$ and $R$ are the spherical moment of inertia, mass and radius
respectively of the source body in the Cavendish experiment, 
\be
U_{\rm ext} = \sum_a \frac{m_a}{r_a} \,, \quad
U_{\rm ext}^{<ij>} =\sum_a \frac{m_a}{r_a} \left ({\hat n}_a^i {\hat n}_a^j -
\frac{1}{3} \delta^{ij} \right ) \,,
\ee
with the sum extending over all masses external to the laboratory, and
where
${\hat e}^i$ and ${\hat n}_a^i$ are
unit vectors pointing from the source body to the test body and to the a-th
external body, respectively.
Angular brackets around the indices denote a symmetric, trace-free
(STF) tensor.  Equation (\ref{anisotropy}) is the special case of Eq.
(\ref{anisotropy2}) for a single external body, and for a point source
mass ($I=0$).  The most
important effect is the anisotropy in $G_{\rm local}$, which can lead to
anomalous Earth tides in geophysics (for the Earth, $I \approx 0.5
MR^2$).  Notice that only the $l=2$, or
quadrupole anisotropy in the external matter distribution contributes.

\subsection{Gravitational radiation reaction}
\label{sec:reaction}

We focus now on the 1.5PN terms in the metric.  Combining the relevant terms
from Eqs. (\ref{hsuba}) with the 1.5PN terms generated by the coordinate
transformations (\ref{transform1}) and (\ref{transform2}), we obtain, 
\bea
h^{(5/2)}_{00} &=& \sum_a  m_a
\left [\frac{8}{3} \dot{\bf a}_a \cdot {\bf z}_a
 - 2 {\bf v}_a \cdot {\bf a}_a + 6 \frac{v_a^2 {\bf v}_a \cdot {\bf
 z}_a}{z_a^2}
  -12 \frac{{\bf v}_a \cdot {\bf z}_a {\bf a}_a\cdot {\bf z}_a}{z_a^2}
   -4\frac{({\bf v}_a \cdot {\bf z}_a)^3}{z_a^4} \right ] 
   \nonumber \\
&& + 4L^0 {\dot U} - 4L^j U_{,j} + 4\Sigma(L^j)_{,j} -4\Sigma(v^j L^0)_{,j}
\,, \nonumber \\
h^{(2)}_{0j} &=& \sum_a m_a
\left [ a_a^j + 4 \frac{{\bf a}_a \cdot {\bf z}_a z_a^j}{z_a^2}
- \frac{v_a^2 z_a^j}{z_a^2} + 2 \frac{({\bf v}_a \cdot {\bf z}_a)^2
z_a^j}{z_a^4
}
- 4 \frac{{\bf v}_a \cdot {\bf z}_a v_a^j}{z_a^2} \right ] 
 \nonumber \\
 && -2 {\dot L}^j + 4L^0 U_{,j} \,,
 \nonumber \\
h^{(3/2)}_{ij} &=& 0 \,,
\label{h15PN}
\eea
where the superscript $(n)$ denotes the order of $\epsilon$.
With these expressions and the geodesic equation,
it is straightforward to derive the 1.5PN contributions to the equation of
motion of a body in the presence of other bodies, 
\be
\frac{dv^j}{dt} = \frac{1}{2} h^{(5/2)}_{00,j} - h^{(2)}_{0j,0} 
- h^{(2)}_{0[j,k]} v^k \,.
\label{eom15PN}
\ee
We restrict attention to a binary system, and evaluate the terms in Eq.
(\ref{eom15PN}) at body 1
(as usual, dropping contributions to potentials due to body 1 itself). 
We use the fact that, at body 1, $L^0 = m_2 \ln r$, $L^0_{,j} = m_2
x^j/r^2$, $L^j = m_2 v_2^j \ln r$, ${\dot L}^j = m_2 a_2^j \ln r - m_2 v_2^j
{\bf v_2} \cdot {\bf x}/r^2$, and so on, where now $x^j = x_1^j-x_2^j$ and
$r=|{\bf x}_1-{\bf x}_2|$; we also recall that $\sum_a m_a a_a^j =0$ from
conservation of momentum at Newtonian order.  The surprising result is
that, despite many cancellations,
there is a residual acceleration at 1.5PN order, given by
\be
a_1^j = 8m_1m_2 \frac{{\dot r} x^j}{r^4} \,.
\ee
The acceleration for body 2 is found by interchanging $m_1$ and $m_2$ and
letting $x^j \to -x^j$.  The relative acceleration $a^j = a_1^j-a_2^j$ 
is then given by
\be
a^j = 16\mu m \frac{{\dot r} x^j}{r^4} \,,
\ee
where $m=m_1+m_2$ and 
$\mu=m_1m_2/m$
are the
total and reduced mass of the system, respectively.
This radiation reaction term does not affect the orbital angular
momentum, but it does cause an {\it increase} in the orbital energy at
the rate $dE/dt= 16\mu^2 m {\dot r}^2/r^3 $. 

We will see in Sec. \ref{sec:binary} that this has disastrous consequences for
Whitehead's theory.

\subsection{Failure of momentum conservation}
\label{sec:momentum}

In gravitational theories that lack suitable conservation laws for
total momentum of gravitating systems, a binary system could suffer
an anomalous acceleration of its center of mass, given in the PPN
framework by
\be
A_{\rm CM} = \frac{1}{2} (\zeta_2 +\alpha_3) \frac{m}{a^2}
\frac{\mu}{m} \frac{\delta m}{m} \frac{e}{(1-e^2)^{3/2}} \,,
\label{cmacc}
\ee
where $\delta m=m_1-m_2$ and the acceleration is directed toward the
pericenter of the lighter body.  In GR (and in any theory based on an
invariant action) the effect vanishes, but in Whitehead's theory, it
does not.

\cite{levicivita} once claimed that this center-of-mass effect
occurred in GR, but 
\cite{EddingtonClark} 
spotted his error and
confirmed that it did not.  \cite{Clark} later showed that the
effect did occur in Whitehead's theory, in agreement with Eq.
(\ref{cmacc}).  At the time, of course, there was no hope of detecting
the effect using known binary systems.  However, the binary pulsar
(Sec. \ref{sec:momentum1913}) provides a particularly stringent bound on
this effect.

\section{Experimental tests of Whitehead's theory}
\label{sec:tests}

\subsection{Gravimeter tests of the anisotropy in $G_{\rm local}$}
\label{sec:gravimeter}

If $G_{\rm local}$ is anisotropic because of the presence of an external
mass, then there will be anomalous tides of the
solid Earth, superimposed on the normal luni-solar tides 
(see \cite{nordtvedtwill72,tegp}
for detailed discussion).  The latter are of
typical
amplitude $\Delta g/g \sim 10^{-8}$ (here $g$ is the local acceleration as
measured by a gravimeter). 
If the external body is the sun itself, then $U_{\rm ext} \sim 10^{-8}$,
and the $G$ anisotropy will produce a tidal signal of comparable
amplitude and of the same frequencies as the solar tide.
It is very unlikely that Whitehead's theory would survive
a comparison between the measured solar Earth tide and
standard tidal theory with such a large additional amplitude.
However, the bound one could achieve has
never been investigated in detail, because a cleaner test is provided by
looking at the so-called sidereal tides. 

If the external mass is that of the
galaxy, then $U_{\rm ext} \sim 5 \times 10^{-7}$, and the direction is fixed
in space.  This produces tides at frequencies associated with the
sidereal day rather than the solar day of the solar tide, and these
can be compared with known sidebands
of the coupled lunar and solar tides.  Measurements by \cite{goodkind}
using superconducting gravimeters showed no evidence of anomalies, and
placed the bound on the Whitehead parameter $|\xi| < 10^{-3}$, as compared
with the Whitehead value of unity.  
This improved upon the earlier bounds of \cite{Will}, which were based
on the existing tidal literature.

This was considered a fatal blow to the theory, but it did assume an
amplitude $5 \times 10^{-7}$ for the anisotropic part of the galactic
potential.  That value came from relating the solar system's orbital
velocity in the galaxy to the potential via $v^2 \sim U_{\rm ext}$.  This
was criticized \citep{Mentock} because it concentrated the mass of the galaxy at
the center, whereas we now know that the bulk of the mass of the galaxy is
in a roughly spherical halo of stars and dark matter, substantially larger
in size than the visible Milky Way.  

However it can be shown using a simple density model for the galaxy that the
original estimate for the anomalous tidal amplitude
holds up within a factor of two.  First, we note that  
the ``trace-free'' tensor potential $U^{<ij>} = -X_{,<ij>}$, where $X$ is the
``superpotential'' defined in Eq. (\ref{LX}).  For a spherically symmetric
distribution of matter, $X$ is given by
\be
X = rm(r) + \frac{1}{3r}\int_0^r 4\pi \rho^\prime r^{\prime 4}dr^\prime
  + \frac{1}{3} \int_r^\infty 4\pi \rho^\prime r^{\prime} (r^2+3r^{\prime
  2})dr^\prime \,,
\ee
where $\rho$ is the mass density and $m(r)$ is the mass inside radius $r$.
Then, for spherical symmetry,
\bea
X_{,<ij>} &=& {\hat n}^{<ij>} (d^2 X/dr^2 - r^{-1} dX/dr ) 
\nonumber \\
&=& -{\hat n}^{<ij>} \left [ \frac{m(r)}{r} - \frac{I(r)}{r^3} \right ]
\,,
\label{Xsubjk}
\eea
where $I(r)$ is the spherical moment of inertia inside radius $r$.   For
flat or monotonically decreasing density distributions. the second term is
always smaller than the first.  

To compare with the earlier estimate we consider a specific density
distribution given by $4\pi \rho = \alpha/r_c^2$, for $r<r_c$, and 
$4\pi \rho = \alpha/r^2$, for $r>r_c$, where $r_c$ is a core radius
meant to represent the mass of the inner part of the galaxy, and
$\alpha$ is a parameter.  The $1/r^2$ density distribution is meant to model
the dark matter halo, and to yield  a flat rotation curve for the outer
reaches of the Milky Way, in rough agreement with observations.  By noting
that a circular orbit in a spherical potential satisfies, 
$v^2/r = a_r =m(r)/r^2$, and considering the case 
$r>r_c$, we can fit $\alpha = v^2/(1-2q/3)$, and find that
\be
U_{\rm ext}^{<ij>} = \frac{2}{3}v^2 {\hat n}^{<ij>} \frac{1-q+q^3/5}{1-2q/3}
\,,
\ee
where $q=r_c/r$.  For the case $r<r_c$, a similar calculation gives
\be
U_{\rm ext}^{<ij>} = \frac{2}{5}v^2 {\hat n}^{<ij>} 
\,,
\ee
independent of $r$.  Thus for $v \sim 220$ km/s, we find an amplitude
$ 2 - 3 \times 10^{-7}$, fully consistent with the earlier estimate.
Note from Eq. (\ref{Xsubjk}) that only the matter inside our
radius has an effect on the anisotropy.  Even though the 
galaxy and its halo are
not strictly spherically symmetric, this is unlikely to alter the estimate
significantly.  The only way to suppress this effect is by some
specific, fine-tuned distribution of external matter.

The conclusion stands: Whitehead's theory violates geophysical tide
measurements by about a factor of $500$.

\subsection{Lunar laser ranging and the Nordtvedt effect}
\label{sec:llr}

In many alternative theories of gravity, there is a violation of the
weak equivalence principle for massive, self-gravitating bodies.
Specifically, the passive gravitational mass $m_p$ may differ from the
inertial mass $m_i$ according to
\be
m_p = m_i (1 - \eta |E_g|/m_i ) \,,
\ee
where 
\be
\eta=4\beta-\gamma-3- \frac{10}{3} \xi -\alpha_1 + \frac{2}{3}
\alpha_2 -\frac{2}{3} \zeta_1 - \frac{1}{3} \zeta_2 \,,
\ee
and $E_g$ is the gravitational binding energy of the body.
This is known as the Nordtvedt
effect \citep{nordtvedt1}, and can cause a difference in acceleration of
the Earth and the Moon toward the Sun, and a resulting perturbation of
the Earth-Moon orbit with a specific signature.  Over 35 years of
lunar laser ranging have found no evidence for such an effect, and
have placed the bound $|\eta| < 9 \times 10^{-4}$ \citep{Williams}.  
From the set of PPN parameter
values for Whitehead's theory in Eq. (\ref{ppnvalues}), $\eta_{\rm
Whitehead} = -1/3$ in strong disagreement with experiment.

\subsection{The binary pulsar}
\label{sec:binary}

Thirty years of timing of the binary pulsar 1913+16 have shown that
its orbital period is decreasing at a rate ${\dot P}_b = - (2.4184 \pm
0.0009) \times
10^{-12}$, in agreement with the GR prediction for gravitational
radiation damping within a fraction of a percent \citep{Weisberg}.  
Orbital damping has
also been measured in the binary
pulsars 1534+12 and the double pulsar 0737-3039AB, again in agreement with
GR.  Unfortunately, Whitehead's theory has both the wrong sign --
antidamping instead of damping -- and the wrong magnitude, ${\dot P}_b
\approx + 4 \times 10^{-8}$.  
The magnitude is so large because the reaction is a 1.5PN 
effect, rather that a $v^2$-times smaller 2.5PN effect, as in GR.
One could change the sign of the effect, but not its magnitude,
by assuming advanced, rather
than retarded interactions.  

\subsection{LAGEOS data}
\label{sec:lageos}

Since 1992, precise laser tracking of two Earth-orbiting 
Laser Geodynamics Satellites
(LAGEOS I and II) has made possible tests of general relativity in the 
vicinity of the Earth, in addition to its primary geophysical goals.
Notably, the tracking data have been used to give a preliminary test
of the ``dragging of inertial frames'', or Lense-Thirring effect, in 
which the rotating Earth
causes a small precession of the planes of the orbits of the
satellites.  The NASA-Stanford Gravity Probe B experiment also aims to
measure this effect with higher accuracy using orbiting
superconducting gyroscopes.  The effect depends on the PPN parameters
$\gamma$ and $\alpha_1$, so both Whitehead's theory and GR agree on
the prediction for this effect.  However, the orbit of the
LAGEOS II satellite has
a small eccentricity, unlike LAGEOS I, and so its advance of perigee
is also measured, along with the ``nodal'' precession of the orbit plane.

Now, the multipole moments of the Earth's Newtonian gravity field also
contribute to the nodal precessions and the perigee advance, indeed
they overwhelm the relativistic effects.  However, 
\cite{Ciufolini97} 
found a particular linear combination of the three measurables,
the two nodal precessions, ${\dot \Omega}_I$, and ${\dot
\Omega}_{II}$, and the perigee precession of II, ${\dot \omega}_{II}$,
in which the effects of the leading $l=2$ and $l=4$ Newtonian
multipoles would precisely cancel.  The combinations depend on the
known inclinations of the orbits relative to the equator.  The
uncertainties in the measured values of the remaining $l \ge 6$
multipoles then become part of the error budget of the experiment.

The only difference in any of the relevant predictions between
Whitehead and GR is the additional size-dependent term in the
pericenter advance, Eq. (\ref{pericenter}).  Because the LAGEOS II satellite is
at two Earth radii, this can be a sizable effect (unlike the case with
Mercury).  Thus, the specific
linear combination of predicted effects used by Ciufolini {\it et al.}
gives the theoretical prediction (in milliarcseconds per year)
\bea
A_{\rm theory} &=&  {\dot \Omega}_I + 0.295 {\dot \Omega}_{II} -0.35 
{\dot \omega}_{II}  \nonumber \\
&=& 60.2 - 109 \xi + ({\rm errors})
\eea
where we have kept the PPN Whitehead parameter $\xi$ but used the
GR/Whitehead values for $\gamma$, $\beta$ and $\alpha_1$, and where
``error'' denotes those due to the higher multipole moments.
Using the actual tracking data, the measured value of this combination
is $A_{\rm exp} = 66.6$ milliarcseconds per year, 
plus measurement errors.  The combination of
all the errors leads to a total estimated error of about 25 percent.  Thus for
the theory to match observation within 25 percent, the parameter $\xi$
must lie in the range
\be
-0.2 < \xi < 0.1 \,,
\ee
which thus excludes Whitehead's theory.
It is likely that this bound could be improved by making use of
dramatically improved Earth gravity models that have been derived from
the GRACE and CHAMP geodesy space missions, which have reduced the
errors in the Earth's multipole moments by significant amounts.

\subsection{Binary pulsars and momentum conservation}
\label{sec:momentum1913}

The binary pulsar B1913+16 provides an excellent system to test the
momentum non-conserving effect described in Sec.~\ref{sec:momentum},
because it is highly relativistic, and because of the ability to do
precise timing.  For a moving system all measured periods will be
offset via the Doppler effect ($\Delta P/P \sim v/c$); 
accordingly, in an accelerating system
periods will suffer a drift $dP/dt \sim (a/c)P$, and in a system 
with a changing acceleration, there will be a $d^2 P/dt^2 \sim
({\dot a}/c)P$.  In the binary pulsar, the center of mass acceleration
predicted by Whitehead's theory changes because it is directed toward
the periastron of the system, which rotates by 4 degrees per year.
Indeed, in the 30 years since discovery, the center-of-mass motion
(were it to exist) would have almost reversed itself.  Yet precise
timing of the pulsar 1913+16 has shown no evidence of any change in
its spindown rate $dP/dt$, leading to an upper bound $|d^2 P/dt^2| <
8.5 \times 10^{-32}$ s$^{-1}$ \citep{manchester}.  
Using the neutron star masses  and orbital elements inferred from
the timing data, together with Eq. (4) of \cite{will92}, we find the
predicted value 
$d^2 P/dt^2 \simeq 2.1 \times 10^{-25}\zeta_2 \cos \omega$ s$^{-1}$, where
$\omega$ is the periastron angle (we adopt the Whitehead value
$\alpha_3=0$).  
With
$\cos \omega$ varying between $-1$ and $+0.59$ over that period, we find
the
bound $|\zeta_2 | < 8 \times 10^{-7}$.  Notice that the
mass values used were inferred using GR; 
in Whitehead's theory, it is conceivable
that these values could be different from the GR values (as occurs in other
theories that violate the strong equivalence principle).  However, to evade
this bound, either the inferred masses would have to be $10^6$ times 
smaller, or they would have to be the same to a part in $10^6$.  
This seems highly unlikely.

\section{Cosmological Considerations} 

In addition to passing  stringent tests at 
terrestial, solar system, and  galactic scales, in order to be viable,
a theory of spacetime and gravity must agree
with the basic facts of cosmology: the expansion of the universe
and the existence of the Cosmic Microwave Background. Of course
neither 
was known when  Whitehead formulated his theory. However at present,
we are entering an era in which  
cosmological observations are becoming increasingly detailed and precise 
\citep{WMAP1}. 

Already during the 1950's \cite{Synge2},
using the spherically symmetric 
continuum version of Whitehead's theory developed
by \cite{Rayner1}, derived the form
the Friedmann-Lemaitre metric takes according to Whitehead.
If $\tau = \sqrt{-\eta_{\mu \nu}  x^\mu x^\nu }$,  the curved  metric is  
\be
ds ^2 = -\left( 1- \frac{3 A}{\tau}\right ) 
d \tau ^2 + \tau ^2 \left ( 1+ \frac{A}{\tau} \right ) 
\bigl ( d \chi ^2 + \sinh \chi  
(d \theta ^2 + \sin ^2 \theta d \phi ^2 ) \bigr )  \,, 
\ee    
with the density
\be
\rho = \frac{K}{\tau ^2} \,, 
\ee
and $A= 8 \pi  G K /9$, with $K$ a constant.
Note that Synge's version of Whitehead's  Universe,
which has hyperbolic, $k=-1$ spatial cross sections, becomes
empty and flat at late times, becoming more and more Milne-like. 

By contrast, current observations \citep{WMAP1}  strongly
indicate that our universe
is currently of Friedmann-Lemaitre  form with flat
spatial sections  and
scale factor $a(\tau)$ with jerk 
\citep{Blandford}
\be
j= { a^2 \over {\dot a}^3 } { d^3 a \over d \tau ^3 }  =1 \,,
\ee 
and thus given by
\be
a(\tau)=  \sinh ^{ 2 /3} 
[  (3 \Lambda /4)^{1/2} \tau ]\,,
\ee 
where $\Lambda$ is the cosmological constant.
As proper time $\tau$ goes by, the universe is more
and more accurately De-Sitter like, with
\be
a(\tau) = e^{(\Lambda/3)^{1/2} \tau} \,.\label{sitter}
\ee 

It seems that to be viable, Whitehead's theory
requires, at the very least, a modification that  incorporates 
the same effects as the  cosmological
term in Einstein's theory.
The principal  motivation behind Whitehead's 
alternative to Einstein's theory
was the desire to retain  fixed, non-dynamical, 
background-independent,  causal 
relations between spacetime events which do not depend upon 
one's location in spacetime.  Presumably, purely on the same
aesthetic or philosophical
grounds,
one might argue that, as a fixed set of
spatio-temporal relations, those of De-Sitter spacetime or
of anti-De-Sitter
spacetime are to be preferred to those of Minkowski spacetime
since the underlying isometry groups in the former two cases
are simple, rather than being a mere  semi-direct product
in the latter.
Be that as it may, early on, \cite{Temple} pointed 
out that this aim could just as readily be achieved 
by adopting the causal relations of a fixed maximally symmetric
spacetime of constant
curvature, e.g. a De-Sitter spacetime,  as it could
by insisting that they were the same as Minkowski spacetime.    
With this in mind, Temple sketched a generalisation
of Whitehead's theory to incorporate a De-Sitter background which
received enthusiastic support from Whitehead himself.
An interesting Machian argument in its favour was made by 
\cite{Band1}, 
who pointed out that for positive cosmological constant
it described a finite universe relative to which one could
define an  absolute acceleration.
Actually Band claimed \citep{Band2} that Whitehead's theory was in gross
violation of experiment.  
Later, \cite{Rayner3} pointed out what he claimed were  some errors
in Temple's formulae. 

Rather than recall the details of Temple's construction,
which appears to have been almost completely forgotten, perhaps
because the reference to it in Synge's influential
reformulation of Whitehead's theory in modern notation
\citep{Synge1} is incorrect,
we shall content ourselves with the remark that the obvious 
statement of the theory\footnote{which however appers to differ
in detail from the approach advocated in \cite{Temple,Rayner3} }
is that it amounts to linearising  
Einstein's theory with a cosmological constant around a 
De-Sitter background. This interpretation is consistent
with Temple's finding that the perihelion advance
agrees with that obtained by Eddington for the Schwarzschild-De-Sitter
metric.  If one accepts our interpretation, then        
the fact that the Kerr-De-Sitter solution is also  of
Kerr-Schild  form \citep{Carter}
shows that Eddington and Rayner's  observations
\citep{Eddington,Rayner2} may  be extended to the 
full set of rotating solutions  in a background
De-Sitter spacetime.  

However, although incorporating a cosmological term may conceivably render
Whitehead's theory in better accord with cosmological data, it will
do nothing to alter the fact that it is in flagrant 
contradiction with  observations at solar system and galactic scales,
since the effects of any cosmological 
modification at these scales are negliglible.

\section*{Acknowledgments}

This work is supported in part by the National Science Foundation under
grant no. PHY03-53180.
CMW is grateful to the Institut d'Astrophysique de Paris 
and to the Institut Henri Poincar\'e,
for their
hospitality while this work was being completed.

\end{document}